\documentclass[aps,prb,showpacs, groupedaddress,twocolumn, superscriptaddress]{revtex4}
\usepackage{amsmath,graphicx,latexsym,times,color}
\usepackage{setspace}
\usepackage{hyperref}
\usepackage{array}
\usepackage{titlesec}

\begin{document}

\title{ Out- versus in-plane magnetic anisotropy of free Fe and Co nanocrystals: tight-binding and first-principles studies }

\author{Dongzhe Li}
\affiliation{CEA, IRAMIS, SPEC, CNRS URA 2464, F-91191 Gif-sur-Yvette Cedex, France}

\author{Cyrille Barreteau}
\affiliation{CEA, IRAMIS, SPEC, CNRS URA 2464, F-91191 Gif-sur-Yvette Cedex, France}
\affiliation{DTU NANOTECH, Technical University of Denmark, {\O}rsteds Plads 344, DK-2800 Kgs. Lyngby, Denmark}

\author{Martin R. Castell}
\affiliation{Department of Materials, University of Oxford, Parks Road, Oxford OX1 3PH, UK.}

\author{Fabien Silly}
\affiliation{CEA, IRAMIS, SPEC, CNRS URA 2464, F-91191 Gif-sur-Yvette Cedex, France}
\affiliation{Department of Materials, University of Oxford, Parks Road, Oxford OX1 3PH, UK.}

\author{Alexander Smogunov}
\email{alexander.smogunov@cea.fr}
\affiliation{CEA, IRAMIS, SPEC, CNRS URA 2464, F-91191 Gif-sur-Yvette Cedex, France}
\date{\today}

\begin{abstract}
We report tight-binding (TB) and Density Function Theory (DFT) calculations of magnetocrystalline anisotropy energy (MAE) of free Fe (body centerd cubic) and Co (face centered cubic) slabs and nanocrystals. The nanocrystals are truncated square pyramids which can be obtained experimentally by deposition of metal on a SrTiO$_3$(001) substrate. For both elements our local analysis shows that the total MAE of the nanocrystals is largely dominated by the contribution of $(001)$ facets. However,  while the easy axis of Fe$(001)$ is out-of-plane, it is in-plane for Co$(001)$. This has direct consequences on the magnetic reversal mechanism of the nanocrystals. Indeed, the very high uniaxial anisotropy of Fe nanocrystals makes them a much better potential candidate for magnetic storage devices.
\end{abstract}



\newcommand{\Ea}{\ensuremath{E_a}}
\newcommand{\Eb}{\ensuremath{E_b}}
\newcommand{\Ec}{\ensuremath{E_c}}
\newcommand{\Ed}{\ensuremath{E_d}}
\newcommand{\Ee}{\ensuremath{E_e}}

\newcommand{\eV}{\ensuremath{\,eV}}

\maketitle

\section{ Introduction }
\label{Sec: Intro}

Last decades, higher storage densities were achieved by reducing the magnetic grains down to nanoscale. However, the magnetic stability of a nano-object decreases proportionally to its size and the ultimate limit is reached when the thermal fluctuation overcomes the energy barrier to swicth the global magnetization of the system. The most crucial issue in exploring ultimate density data storage (e.g., high-density magnetic recording \cite{Albrecht_2002} or spintronic devices) is magnetic anisotropy energy, which is defined as the change of total energy associated to a change in the direction of magnetization.  One of the challenge in this route towards high magnetic density storage is evidently to be able to synthetize well ordered arrays of magnetic nanocrystals with as large magnetization and magneto-crystalline anisotropy as possible. The magnetic anisotropy energy of magnetic nanocrystals (e.g., Fe, Co and Ni etc) is indeed a subject of intense study both experimentally \cite{Balashov_2009, Jamet_2004, Jamet_2001, Luis_2002} and theoretically \cite{Pastor1995, Nicolas2006,Xie_2006, Lazarovits_2002,Medina_2003, Sahoo_2010} but the ability to  grow well defined magnetic crytalline nanostructures is also a major issue \cite{Kim_2002, Qiang_2002, Martina_2003, Silly2005_Co, Silly2005_Fe, Sun_2013, Ding_2005}.
This is especially the case for Fe and Co nanostructures, that can adopt various crystalline bulk structures, in particular the body-centered cubic (bcc) and face-centered cubic structure (fcc) structure in low dimensions \cite{ Silly2005_Fe, Silly2005_Co}. The close-packed and lowest-energy facet for bcc structure is the (110) facet whereas it is the (111) facet for the fcc structure. This is the reason for the (110) facets appearing in bcc nanocrystals (in fact for Fe the surface energies of $(001)$ and $(110)$  orientations are almost the same)  and the (111) facets appearing in fcc nanocrystals. The nanocrystal magnetic properties will therefore not only depend on the bulk structure but also on the facet structure and area.

The magnetic anisotropy contains two different parts:  the first part is long-range magnetic dipole-dipole interaction which leads to so-called shape anisotropy, while the second one is referred to magnetocrystalline anisotropy energy (MAE) originates from the spin-orbit coupling (SOC) \cite{Bruno1989}. The latter effect is a quantum effect (of relativistic nature) that breaks the rotational invariance with respect to the spin quantization axis. Therefore, if SOC is included, the energy of the system  depends on the orientation of the spin with respect to the crystallographic axis.

The value of MAE per atom, is extremely small in bulk (some $\mu$eV), but can get much larger in nanostructures \cite {Gambardella_2003, Rusponi_2003} (some meV) due to reduced dimensionality. From the point of view of theory, there are two different methods which are used extensively in the literatures for MAE calculations: i) fully relativistic self-consistent field (SCF) calculations, ii) Force Theorem (FT) \cite{Wang1996, Daalderop1990,Dongzhe_PRB}. Assessing the MAE for systems containing hundreds of atoms by the former approach is especially challenging, since it requires a well-converged charge density as well as a  consuming computational time. In the latter method, the MAE is given by the band energy difference (instead of total energy difference) obtained after a one-step diagonalization of the full Hamiltonian including SOC, starting from the self-consistent scalar relativistic density/potential. This approach is not only computationally efficient but also numerically very stable since the self-consistent effect with SOC could be ignored.

The objective of this paper is to investigate the MAE of Fe and Co nanocrystals, that can be grown experimentally by epitaxy, 
 using tight-binding (TB) as well as first principles calculations in the Density Functional Theory (DFT) framework. The nanocrystals adopt a truncated-pyramid shape on a reconstructed SrTiO$_3$(001) substrate but have however a different bulk structure (bcc and fcc). In our theoretical study a particular emphasis is devoted to the local analysis of MAE. In particular, it has been found that the main contribution to the MAE comes from the basal (001) facet of pyramids. This results in strong out-of-plane (in-plane) anisotropy for Fe (Co) nanocrystals, in agreement with the study of thick Fe(001) and Co(001) slabs.

The paper is organized as follows. In Sec. \ref{Sec: Methodolody} we present the experimental and theoretical methods used in this work. In Sec. \ref{Sec: Results}, we first present Scanning Tunneling Microscope (STM) observation of Co nanocrystals on SrTiO$_3$(001) substrate and illustrate the results of TB and DFT calculations for Co(001) and Co(111) slabs. 
After that, the MAE of free Fe and Co nanocrystals will be discussed. Finally, the conclusions will be presented in Sec. \ref{Sec:conclusion}.

\section{ Methodolody }
\label{Sec: Methodolody}
In the following sections we will first briefly present the main ingredients of the experimental set-up to grow the cobalt nanocrystals on the SrTiO$_3$(001) substrate. Then in the second part the theoretical model to calculate the MAE will be presented starting by the TB Hamiltonian and then the DFT approach. In the case of the DFT formalism we will essentially concentrate on the implementation of the Force Theorem which we have incorporated in the Quantum-ESPRESSO (QE) \cite{Giannozzi2009} package.

\subsection{ Experimental }
\label{Sec:STM_model}
we use SrTiO$_3$(001) crystals doped with 0.5\% (weight) Nb. The crystals were epi-polished (001) and supplied by PI-KEM, Surrey, UK. We deposited Co from an e-beam evaporator (Oxford Applied Research EGN4) using 99.95\% pure Co rods supplied by
Goodfellow, UK. Our STM is manufactured by JEOL (JSTM 4500s) and operates in ultra high vacuum (10$^{-8}$ Pa). We used etched W tips for imaging the samples at room temperature with a bias voltage applied to the sample. SrTiO$_3$(001)-c(4$\times$2) was
obtained after Ar$^{+}$ bombardment and annealing in UHV at $600\,^{\circ}\mathrm{C}$ for 2 hours. STM images were processed and analyzed using the home made FabViewer application \cite{Fabien_2009}.

\subsection{ Theoretical }

\subsubsection{ Magnetic tight-binding model }
\label{Sec:TB_model}
In this section, we briefly describe our magnetic tight-binding model (more details can be found in our previous publications \cite{Autes2006, Barreteau2012}). The hamiltonian is written as follows:
\begin{equation}
H=H_{\text{TB}}+H_{\text{LCN}}+H_{\text{Stoner}}+H_{\text{SOC}}
\end{equation}
Where $H_{\text{TB}}$ is a standard "non-magnetic" TB hamiltonian which form is very similar to the one introduced by Mehl and Papaconstantopoulos\cite{Mehl1996}, $H_{\text{LCN}}$ is the term ensuring a local charge neutrality, $H_{\text{Stoner}}$  is the Stoner-like contribution that controls the spin magnetization and $H_{\text{SOC}}$ corresponds to spin-orbit coupling that operates on $d$ orbitals only.

The total energy should be corrected by a double counting term due to inter-electronic interactions introduced by local charge neutrality and Stoner interaction, explicitly:
\begin{equation}
\begin{split}
E_{\text{tot}}=E_{\text{b}}-E_{\text{dc}}=\sum_{\alpha} f_{\alpha} \epsilon_{\alpha}-
\frac{U}{2}\sum_i [N_i^2-(N_i^0)^2]+ \\ 
\frac{1}{4}\sum_{i,\lambda} I_{\lambda}M_{i \lambda}^2,
\end{split}
\end{equation}
where $E_{\text{b}}=\sum_{\alpha} f_{\alpha} \epsilon_{\alpha}$ is the band energy, $f_{\alpha}$ is the Fermi-Dirac occupation of state $\alpha$,  $N_i$ and $M_i$ are the charge and the spin moment of site $i$, respectively. $N_i^0$ is the valence charge, $U$ is the parameter 
imposing the local charge neutrality, and $I_\lambda$ is the Stoner parameter of the orbital $\lambda$  ($\lambda=s,p,d$).

All the parameters of TB hamiltonian are fitted on bulk {\it ab initio} data: bandstructure, total energy, magnetic moment etc. The value of the Stoner parameter $I_d$  is taken equal to 0.88 meV for Fe and 1.10 meV for Co. The spin-orbit constant $\xi_d$ is also determined by comparison with {\it ab initio} bandstructure and we found that 60 meV and 80 meV are very good estimates for Fe and Co, respectively.

The MAE, in  a very good approximation, is calculated by using the Force Theorem (FT) \cite{Wang1996, Daalderop1990, Dongzhe_PRB}:  
first, a self-consistent field (SCF) collinear calculation without SOC is done followed by the rotation of the density matrix in the right spin direction; next,  a non-SCF non-collinear calculation with SOC is performed. The MAE is obtained as the difference of band energies, 
$E_{\text{b}}^1-E_{\text{b}}^2$, between two spin moment directions, 1 and 2. The correct decomposition of total MAE over different atomic sites $i$ can be done within so-called "grand canonical" formulation \cite{Dongzhe_PRB}:
\begin{equation} \label{FTgc_atomic}
MAE_i=\int\limits^{E_{F}}(E-E_{F})\Delta n_{i}(E)dE
\end{equation}
where $\Delta n_{i}(E)=n_{i}^1(E)-n_{i}^2(E)$ is the change in the density of states at atom $i$ for different spin 
moment orientations. The Fermi energy $E_{\text{F}}$ of SCF calculation without SOC
must be subtracted from all energies in order to suppress the trivial contribution to the local MAE
due to charge redistribution as discussed in Ref.\cite{Dongzhe_PRB}.
 
\subsubsection{ Density Functional Theory (DFT) calculations }
\label{Sec:DFT_model}
We perform {\it ab initio} DFT calculations using the plane-wave electronic structure package
Quantum-ESPRESSO (QE).\cite{Giannozzi2009} The spin-orbit coupling (SOC), crucial for magnetocrystalline anisotropy, 
is taken into account via fully-relativistic pseudo-potentials (FR-PPs)
\cite {Corso_2005}, describing the interaction of valence electrons with ions,
which are in turn generated by solving atomic Dirac equations for each atomic type. 
We have implemented the Force Theorem in QE in the same two-step way as described above for TB model:
i) SCF calculation with scalar-relativistic PPs (without SOC) is performed to obtain the 
charge density and the spin moment distributions in real space;  ii) spin moment is globally rotated 
to a certain direction followed by a non-SCF calculation with FR-PPs (with SOC). 
The change of band energy between two spin moment directions gives, as above, the total MAE.

The total MAE is decomposed over different atoms $i$ in the slightly different way:     
\begin{equation}
MAE_i=\int\limits^{E_F^1}  (E - E_F^2)n_i^1(E)dE-\int\limits^{E_F^2} (E - E_F^2)n_i^2(E)dE,
\end{equation}
where the Fermi level of one of magnetic configurations 
(we have chosen the second one),  $E_{\text{F}}^2$, is substracted under integrals 
and exact Fermi levels for two configurations are used as the limits of integration.
This way we avoid the reference to electronic levels of a system without SOC, since the PPs with and without SOC
are not generally correlated and can produce an arbitrary shift of levels.
Due to total charge conservation in this "canonical" approach, the sum of $MAE_i$ over all
atoms gives exactly the total MAE while for the "grand canonical" scheme, Eq. \ref{FTgc_atomic}, it was, in principle, only approximate.
The descrepancy between "grand canonical" and "canonical" formulations within TB approach is, however, very tiny
since the effect of SOC on the Fermi level is negligable in the case of Fe or Co composed materials.

\begin{figure}[!hbp]
\centering
\includegraphics[scale=0.4]{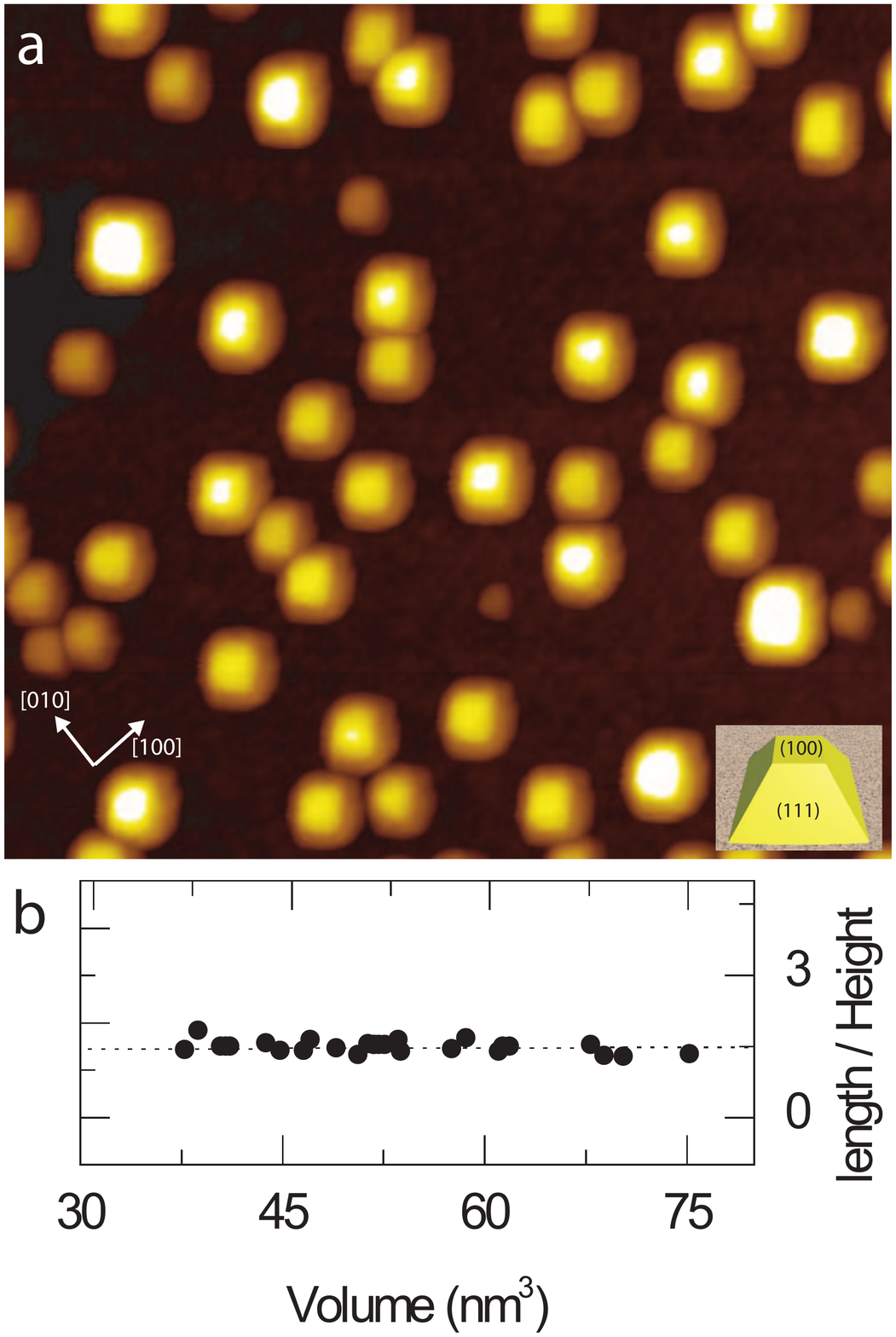}
\caption{\label{fig:figSTM}
(color online) (a) Co deposition onto a $350\,^{\circ}\mathrm{C}$ SrTiO$_3$(001)-c(4$\times$2) substrate followed by a 320$^{\circ}$C anneal gives rise to truncated pyramid shaped nanocrystals as shown in the STM image, 80$\times$80~nm$^2$, V$_s$~=~+1.0~V, I$_t$~=~0.1~nA. (b) height to length ratio constant of Co nanocrystals.}
\end{figure}

Since QE gives an access to real space wave-functions it is natural to define also the space-resolved MAE as:
\begin{equation}
\label{mae_r}
\begin{split}
MAE(r)=\int^{E_F^1}  (E - E_F^2)n_1(r,E)dE- \\
\int^{E_F^2} (E - E_F^2)n_2(r,E)dE,
\end{split}
\end{equation}
where the LDOS is computed via electron wave-functions in the usual way, 
$n^{1,2}(r,E) = \sum_\alpha | \Psi_\alpha^{1,2}(r) |^2\delta(E-\varepsilon_\alpha^{1,2})$. 
Once again, the integral of MAE(r) over all the space will give exactly the total MAE.

\section{ Results and discussion}

In this section we will first briefly present the structural characterization of supported Co nanocrystals using STM. We then present the results of our calculations on slabs of fcc Co with orientations $(001)$ and $(111)$ corresponding to the facets of the nanocrystals. Next, in the second part we consider Fe and Co nanocsrystals in form of truncated pyramids with the same length to height ratio as in the experiments.
\label{Sec: Results}

\subsection{ STM Observations }
\label{Sec:STM_Results}

The SrTiO$_3$(001)-c(4$\times$2) surface \cite{Castell_2002} is used for cobalt deposition. The c(4$\times$2) reconstruction was verified by STM and LEED before deposition. Fig.~\ref{fig:figSTM} shows the topography of the SrTiO$_3$(001)-c(4$\times$2) surface following deposition of 3 monolayers (ML) of Co on a substrate heated to $320\,^{\circ}\mathrm{C}$ followed by a
subsequent 50 minute anneal at $350\,^{\circ}\mathrm{C}$. The Co has self-assembled into similarly sized nanocrystals. Cobalt usually adopt a hcp bulk structure but STM shows that Co nanocrystals have a fcc structure with  the shape of a truncated pyramid (a square top surface and a square base). The Co nanocrystals have. The side facets of the nanocrystals were measured at an angle of $\sim$54$^{\circ}$ with respect to the substrate. This shows that Co is cubic packed and the nanocrystals have a (001) top facet and four (111) side facets. The interface is therefore a (001) plane and the interface crystallography is $(001)_{\text{Co}}$ $\|$ $(001)_{\text{SrTiO$_3$}}$, $[100]_{\text{Co}}$ $\|$ $[100]_{\text{SrTiO$_3$}}$. As a guide to the eye we have shown in Fig.~\ref{fig:figSTM} (inset) a schematic illustration of a truncated pyramid. The ratio of the length ($\ell$) of the top square to the height ($h$) of the truncated pyramids as a function of volume is shown in Fig.~\ref{fig:figSTM}b. The constant ratio of $\ell/h$=1.48$\pm$0.13 suggests that these pyramidal nanocrystals have reached their equilibrium shape. The error in the ratio denotes the standard deviation of the measurements. 

\subsection{ Calculations }

\begin{figure}[!hbp]
\centering
\includegraphics[scale=0.3]{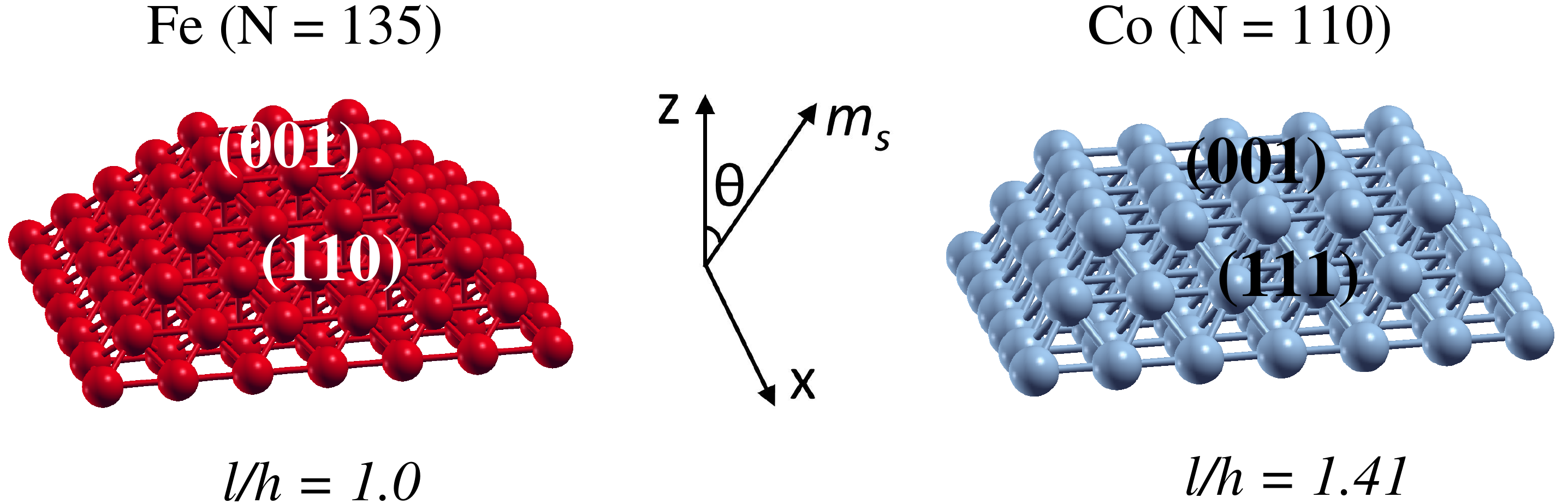}
\caption{\label{pyr_structure}
(color online) Examples of truncated-pyramid shaped Fe and Co nanocrystals studied in the present work. The crystals are made of bcc Fe and fcc Co with two types of facets: (001) and (110) for Fe and (001) and (111) for Co, respectively. Their possible size and shape is controlled by length-to-height ratio, $l$/$h$,  kept to $\sim$ 1.0 (Fe) and 1.41 (Co) which are close to experimental values, $\sim$ 1.20 (Fe) \cite{Silly2005_Fe} and $\sim$ 1.48 (Co). The $z$ axis was chosen to be normal to the pyramid base and the spin moment is rotated in the $xz$ plane forming the angle  
$\theta$ with the $z$ axis.}
\end{figure}

As has been discussed above, fcc Co nanocrystals (Fig.~\ref{fig:figSTM}) as well as  bcc Fe nanocrystals \cite{Silly2005_Fe} can be epitaxially grown 
on SrTiO$_3$$(001)$ substrate with a remarkable control of size, shape and  structure. These crystals can contain up to several hundreds of atoms and have the form of truncated pyramids, as shown in Fig.~\ref{pyr_structure}, with a rather constant length-to-height ratio, $l/h$. They however adopt different bulk structure, i.e. the nanocrystal facets will therefore be different because the close-packed and lowest-energy facet for bcc structure is the (110) facet whereas it is the (111) facet for the fcc structure. It is expected that the MAE of such pyramids will be dominated by the surface composed of (001) and (110) or (001) and (111) facets for Fe and Co nanocrystals, respectively. It is therefore essential to estimate first the MAE of the bulk slabs of these orientations. We present below the results for fcc Co $(001)$ and $(111)$ slabs while similar results for bcc Fe slabs have already been reported recently (Ref. \cite{Dongzhe_PRB}). 

\subsubsection{ MAE of Co fcc $(001)$ and $(111)$ slabs }
\label{Sec:slabs}
\begin{figure}[!hbp]
\centering
\includegraphics[scale=0.5]{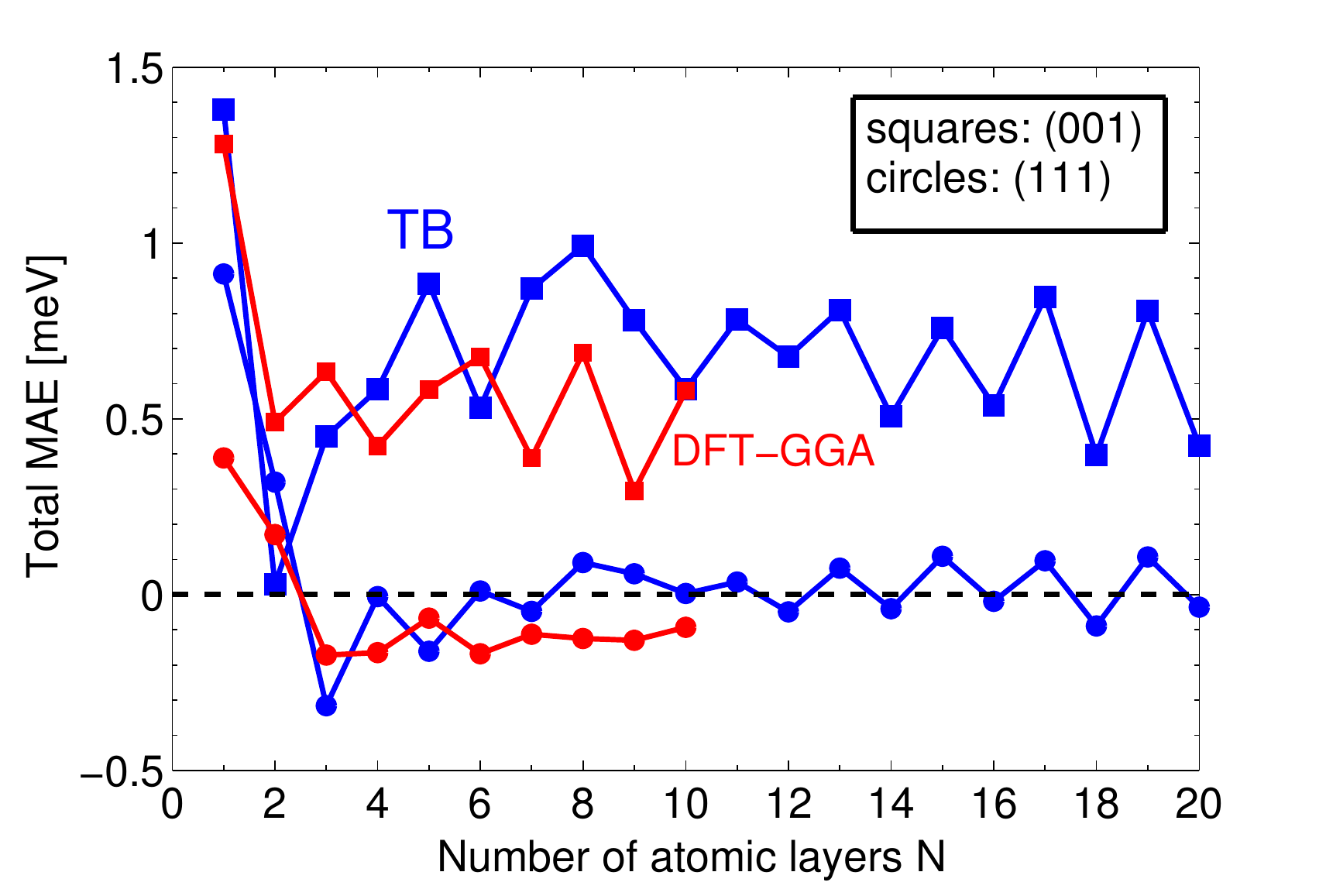}
\caption{\label{slab_total}
(color online) Total MAE per unit cell of N atoms (in meV), namely, $E^{\perp}_{\text{tot}}$ - $E^{\parallel}_{\text{tot}}$, versus the Co film thickness N for fcc Co slabs. Squares and circles are for (001) and (111) slabs respectively. TB calculations (blue) are compared with {\it ab initio} DFT-GGA fully relativistic calculations (red) until N = 10. Lines are guide for the eyes.}
\end{figure}

The Co slabs were constructed from fcc Co with a lattice parameter of $a_0=3.531$~\AA~ found from 
{\it ab initio} calculations (which is close to the experimental value of $a_0=3.548$~\AA~)  and no atomic relaxations were performed. Fig. \ref{slab_total} shows thickness dependence of the total MAE of N-atom fcc Co slabs of (001) and (111) orientations. The results of both TB (N = 1$\sim$20) as well as {\it ab initio} (N = 1$\sim$10) calculations are presented. Note that the total MAE is obtained as total energy difference for $\vec{M}$ perpendicular or parallel to the atomic slabs. Explicitely,  
\begin{math}
\Delta E=E^{\perp}_{\text{tot}}-E^{\parallel}_{\text{tot}}
\end{math}.

In the TB model, a mesh of $50 \times 50$ in-plane $k$ points has been used for SCF calculations without SOC whereas the mesh was incresed to $70 \times 70$ in non-SCF calculations with SOC in order to provide a precision below $10^{-5}$ eV. A Marzari-Vanderbilt broadening scheme with smearing parameter of 50 meV has been used. 
{\it Ab initio} DFT calculations were carried out with Quantum ESPRESSO package \cite{Giannozzi2009} using generalized gradient approximation (GGA) for exchange-correlation potential in the Perdew, Burke, and Ernzerhof parametrization \cite{pbe_1996}. Full self-consistent calculations were performed with relativistic ultrasoft pseudo-potentials (no FT was employed here) and  cut-off energies were set to 30 and 300 Ry for wave-functions and charge density, respectively. The mesh of $40 \times 40$ $k$ points was used and the same smearing parameter and technique were employed.

We find a relatively good overall agreement between TB and DFT calculations. MAE oscillations for both slabs can be clearly seen even for quite thick slabs (similar results were recently reported for bcc Fe slabs\cite{Dongzhe_PRB}). This kind of long-range oscillating behavior has been recently reported by experiments in thin ferromagnetic films (Fe and Co), and was interpreted in terms of spin-polarized quantum well sates.\cite{Przybylski2012, Manna2013}. We notice further that for Co(001) slabs both calculations give rather similar results: the total MAE clearly favours in-plane magnetization with anisotropy energy around 0.6 meV/cell. In the case of Co(111), the MAE oscillates around zero in TB model while the DFT calculations predict rather small (compared to the (001) case) out-of-plane magnetic anisotropy. Note that our results compare rather well with DFT calculations in Ref. \cite{Zhang_2009} done with LDA approximation for exchange-correlation functional.
\begin{figure}[!hbp]
\centering
\includegraphics[scale=0.5]{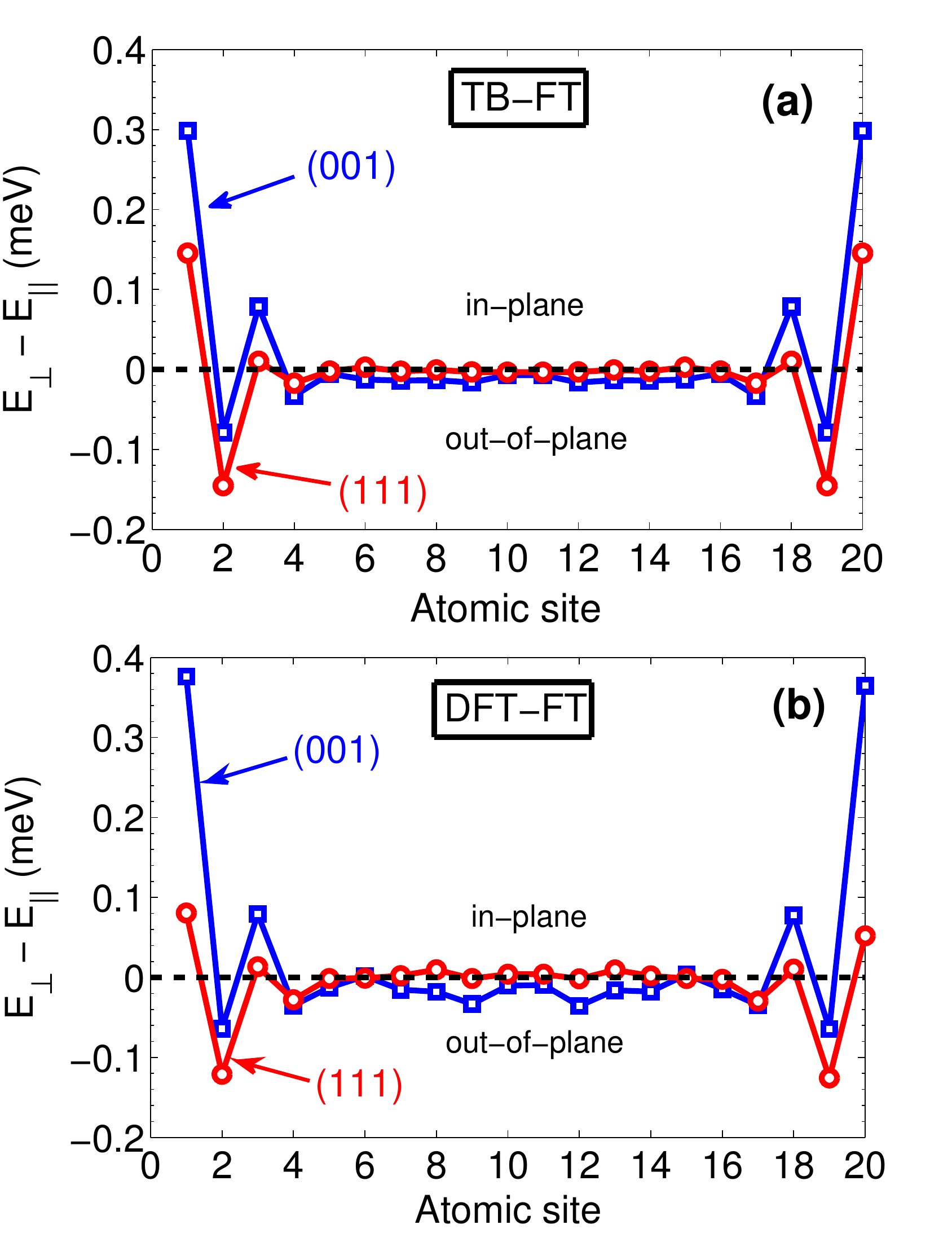}
\caption{\label{slab_local}
(color online) Layer-resolved MAE per Co atom (in meV) of slabs with 20 atomic-layer-thick calculated by Tight-Biding (top) and DFT-GGA (bottom)  within force theorem approximation. Blue squares and red circles are for (001) and (111) slabs, respectively. Lines are guide for the eyes.}
\end{figure}
We further study the local decomposition of MAE of (001) and (111) Co slabs made of 20 atomic layers (Fig. \ref{slab_local}). Here, we used the FT in TB as well as in DFT approaches as described in the previous section. A qualitatively good agreement between TB and DFT calculations is again found for both slabs with the main discrepancy appearing for the surface layers, which indicates that the TB model is presumably less accurate for low coordinated atoms. Interestingly, for both (001) and (111) slabs these surface layers possess in-plane anisotropy. The local MAE site decomposition then shows damped oscillations converging towards a tiny bulk value. However, while the MAE of the (001) slab is strongly dominated by the outermost surfaces layer, this is not the case for the (111) slab where sub-surface layers cancel (and even overcome in the DFT case) the surface contribution. This leads to the large in-plane and rather small out-of-plane overall MAE for the (001) and (111) slabs, respectively, as it is reported in Fig. \ref{slab_total}.   

\subsubsection{ Free Fe and Co nanocrystals }

\label{table_cluster}
\begin{center}
\begin{table*}[ht]
 \begin{tabular}{c|ccc|ccc}
\toprule 
\multicolumn{1}{c|}{} & \multicolumn{3}{c|}{Fe (N=620)} & \multicolumn{3}{c}{Co (N=615)} \\
\multicolumn{1}{c|}{$$} & \multicolumn{1}{c} MAE (meV) & \multicolumn{1}{c} MAE/atom (meV) & \multicolumn{1}{c|} {N atoms} &\multicolumn{1}{c}  MAE (meV) & \multicolumn{1}{c} MAE/atom (meV) &\multicolumn{1}{c}  N atoms \\ \hline
	upper perimeter & -4.203 & -0.262 & 16 & 5.799 & 0.181 &  32\\ 
	upper (001) & -3.541 & -0.393 & 9 & 18.091 & 0.369 & 49 \\ 
	lower perimeter & -37.265 & -0.846 & 44 & 41.590 & 0.866 & 48   \\ 
	lower (001) & -52.839 & -0.528 & 100  & 41.297 & 0.341 & 100 \\ 
    	side surfaces  & -11.457 & -0.063 & 180 & 1.232 & 0.010 & 120 \\
    	total & -103.473 & -0.166 &620  & 112.711 & 0.183 &615 \\ \hline\hline

\end{tabular}
\caption{TB results: Local analysis of MAE for Fe (N=620) and Co (N=615) nanocrystals, note that sign negative (positive) means out-of-plane (in-plane) magnetization.}
\label{table_cluster}
\end{table*}
\end{center}

\begin{figure}[!hbp]
\centering
\includegraphics[scale=0.5]{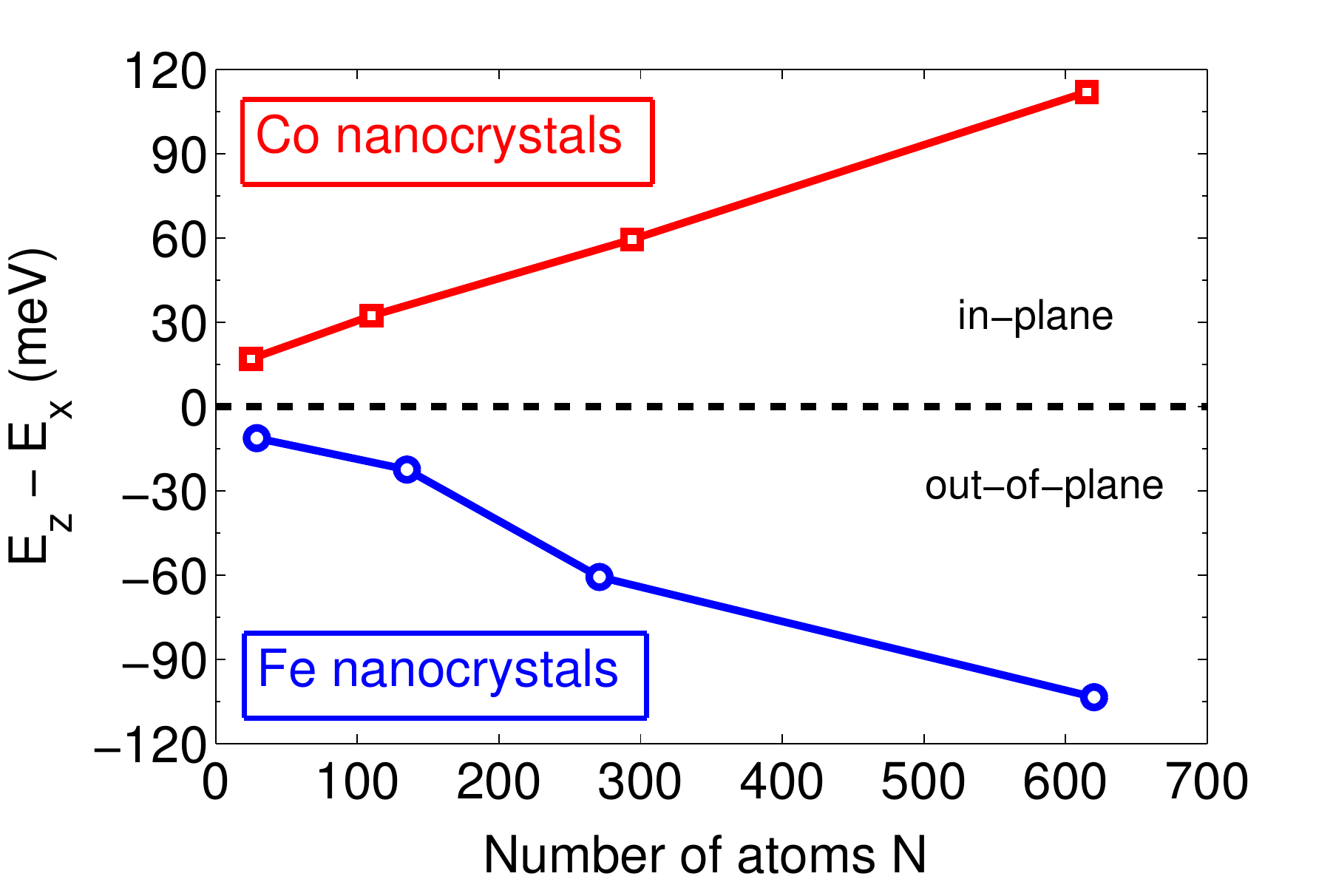}
\caption{\label{pyr_total}
(color online) TB results: total MAE of Co (red squares) and Fe (blue circles) nanocrystals vs. the number of atoms. The size of nanocrystals was chosen so to keep constant length-to-height ratio, 1.41 (Co) and $\sim$1.0 (Fe). }
\end{figure}

The length-to-height ratio of different size of Fe and Co nanocrystals can be written $l/h = [2(n-1)]/(N-n)$ and $l/h = (n-1)/[\sqrt{2}(N-n)]$, where N $\times$ N and n $\times$ n are the number of atoms in the first (bottom) and last (up) layers of the truncated pyramids. We then selected different sizes of  Fe and Co nanocrystals with the length-to-height ratio of $\sim$ 1.0 ($l/h$ = 1.0 for N =29, 135; 1.20 for N = 271; 1.14 for N = 620) and 1.41 close to the experimental value of 1.20$\pm$0.12 \cite{ Silly2005_Fe} and 1.48$\pm$0.13, respectively.   Since the MAE in the $xy$ plane was found  
to be extremely small, we kept the magnetization always in the $xz$ plane making the angle $\theta$ with the $z$ axis.
The MAE is defined as the change in the band energy between magnetic solutions with magnetization along the $z$ and $x$ axis, $MAE=E_z-E_x$. In Fig. \ref{pyr_total}, we plot the total MAE of Fe and Co nanocrystals of growing size calculated with TB approach. Different sign of MAE means that out-of-plane magnetization is favored in Fe while in Co
the spin moment will be rotating in the easy $xy$ plane which makes thus Fe nanocrystals a better candidate for magnetic storage applications. These results can be understood from the local analysis reported in Table \ref{table_cluster} for biggest Fe (N = 620) and Co (N = 615) pyramids.

One can see that the total MAE mainly originates from the lower (001) facet and its perimeter composed of least coordinated atoms. 
Therefore, in agreement with the previous analysis of (001) Co and Fe\cite{Dongzhe_PRB} slabs, this would favor the out-of-plane/in-plane anisotropy for Fe/Co nanocrystals, respectively. We notice, moreover,  that since nanocrystals of Co are much flatter then those of Fe (as Fig. \ref{pyr_structure} illustrates), which is a consequence of  bigger length-to-height ratio for Co,  
in the case of Co nanocrystals also the upper (001) facet, containing more atoms,  gives noticeable contribution to the overall MAE. 
We have also checked the total MAE in the $xy$ plane but have found it extremely small, of amplitude about 3 meV and 
0.8 meV for Fe (N=620) and Co (N=615) nanocrystals, respectively. 
As mentioned in Sec. \ref{Sec: Intro}, another important contribution to magnetic anisotropy is the so-called shape anistropy energy. We have calculated it for biggest Fe (N=620) and Co (N=615) nanocrystals and have found rather small values, of about 5 meV and 2 meV for Fe and Co, respectively. Note that for both pyramids, the shape anisotropy favors in-plane magnetization direction.

\begin{figure}[!hbp]
\centering
\includegraphics[scale=0.55]{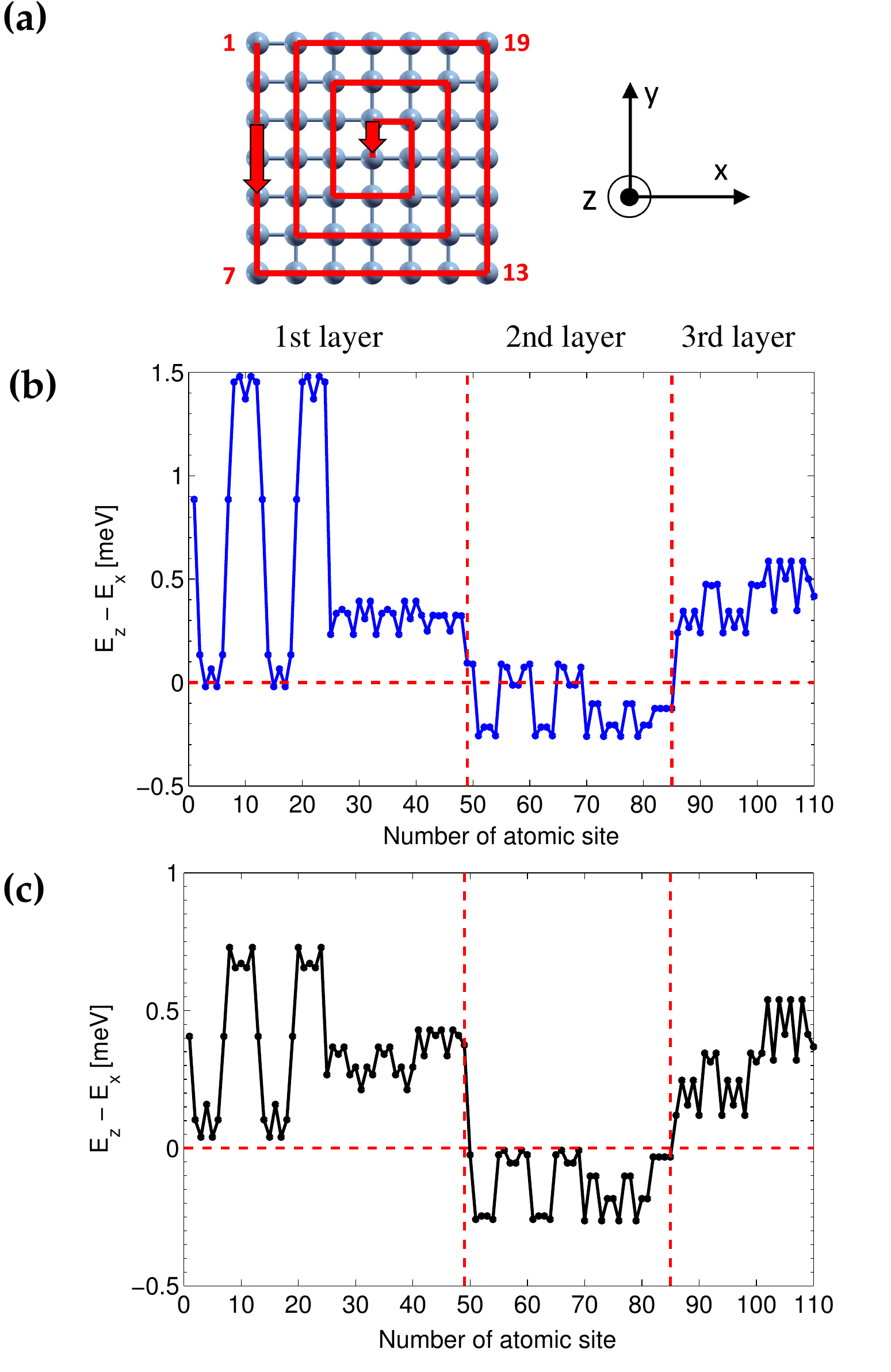}
\caption{\label{pyr_local}
(color online) 
Atom-resolved MAE for Co nanocrystal made of 110 atoms: (a) trajectory for numbering the base layer atoms  starting from the corner and going along the spiral to the center. The atoms of other layers are numbered in the similar way. (b) MAE per atom in meV within TB approach. (c) MAE per atom in meV from DFT-GGA calculations. }
\end{figure}

\begin{figure}[!hbp]
\centering
\includegraphics[scale=0.3]{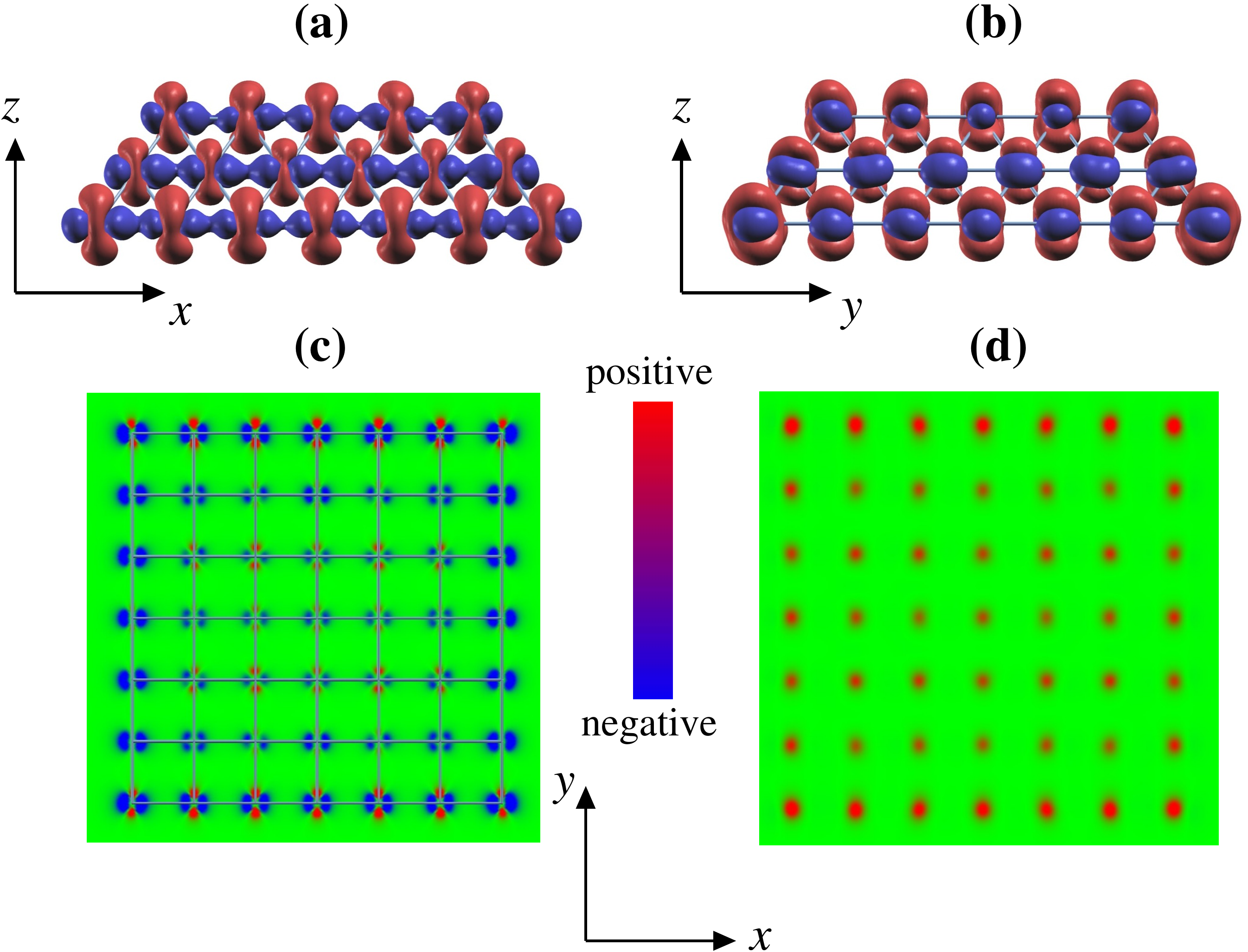}
\caption{\label{pyr_3d}
(color online) DFT calculations: real-space distribution of MAE for Co nanocrystal of 110 atoms: (a),(b) side views, two isosurfaces of positive and negative isovalues are shown in red and blue, respectively;  (c) cross-section of MAE by the plane passing through the base layer of the pyramid; (d) same as (c) but on the plane slightly below (by 0.4~\AA) the base layer. Note that red (blue) colors represent the regions favoring  in-plane (out-of-plane) magnetization orientation.}
\end{figure}

We have next performed a more detailed local analysis of MAE for a smaller Co nanocrystal made of 110 atoms 
(shown on the right panel of Fig. \ref{pyr_structure}). For such a relatively small crystal, {\it ab initio} DFT calculations within FT approach can be also carried out and compared with TB results. Fig. \ref{pyr_3d} reports atom-resolved MAE for such pyramid.
The atoms of each atomic layer are numbered starting from the corner and going anticlockwise along the spiral 
to the centre of the plane, as shown in Fig. \ref{pyr_local} (a) for the base layer. The other layers are numbered in the same way.
Again, a qualitatively good agreement has been found between TB and DFT calculations.
Interestingly, we found a sign change of MAE between atomic layers: the MAE favors in-plane magnetization for the first and third layers and out-of-plane magnetization for the middle layer of the pyramid.
The MAE achieves its highest values in the middle of two first layer edges aligned with the $x$ axis,
namely for 7-13 and 19-1 segments, and drops down to zero for two other edges. This asymmetry is due to chosen 
definition of $MAE=E_z-E_x$, since for the first pair of edges we compare the energies between orthogonal and parallel to the edge
directions while for the second pair -- between two perpendicular directions. Clearly, in the first case the energy difference will be 
much bigger. Of course, if one chooses another definition of MAE, e.g., as the energy difference between the states with spin moment along the $z$ axis and along the diagonal of the base plane, one would have more symmetric contributions from all four base edges.
 
To get more insight into the local composition of MAE, we have looked at its distribution in the real space as defined in Eq. \ref{mae_r}.
Such a real space representation of MAE for the previously studied 
110 atoms Co pyramid is shown in Fig. \ref{pyr_3d}. Interestingly, there are regions of both positive as well as 
negative MAE around each atom (Fig. \ref{pyr_3d} a, b), in relative proportion which changes from layer to layer. This leads, on average, to the change of sign for atomic 
MAE vs. the layer observed in Fig. \ref{pyr_local}. We notice moreover that positive and negative regions of MAE have different spatial localization: 
while the first one extends out of atomic planes (along the $z$ axis) the second one is mostly localized in the $xy$ plane. 
This observations can be important when studying the MAE modification due to deposition of pyramids on various substrates.

\section{Conclusion}
\label{Sec:conclusion}
We have presented a combined TB and DFT study of magnetocrystalline anisotropy of iron (bcc) and cobalt (fcc) slabs and nanocrystals. The nanocrystals are in shape of truncated pyramids with the same length to height ratio as in the experiment. Thanks to the use of the Force Theorem that we have recently implemented in the QE package we have been able to perform a careful local analysis of the MAE in these nanostructures. The TB model is in good agreement with the DFT calculations and gives us confidence in the validity of our TB results for large crystals that cannot be done within the DFT approach.  We found a large in-plane anisotropy for Co$(001)$ and a relatively small out-of-plane for Co$(111)$ due to cancelation from the sub-surface layer in the latter case.  This is in contrast with iron surfaces since Fe$(001)$ shows a clear out of plane anisotropy. The densest surface shows however a rather small anisotropy for both elements. 

These results could have a direct consequence on the magnetic stability of Fe and Co nanocrystals. 
Indeed, the total MAE is of the same order of magnitude for both Fe and Co nanocrystals, but opposite in sign. 
This means that while the spin moment of Fe nanocrystals is fixed along the easy out-of-plane axis and needs to overcome the high MAE
barrier to reverse from positive to negative direction,  
the magnetic moment of Co nanocrystals is allowed to rotate almost freely (with a very low in-plane anisotropy barrier) in the easy 
basal plane.
One can thus conclude that Fe nanoclusters should be better candidates for magnetic storage applications. 
Our local analysis, however, indicates that the MAE of nanocrystals could be substantially altered, for instance, by their covering 
with a mono-layer of another chemical element or by their deposition on various substrates (SrTiO$_3$(001), Cu, Au etc).

\noindent{\bf Acknowledgement}

The research leading to these results has received funding from the European Research Council under the European Union's Seventh Framework Programme (FP7/2007-2013) / ERC grant agreement n$^{\circ}$ 259297. The authors would like to thank the Royal Society and DSTL for funding and Chris Spencer (JEOL UK) for valuable technical support. This work was performed using HPC resources from GENCI-CINES (Grant Nos.  x2013096813).


\bibliographystyle{apsrev}
\bibliography{References}

\end{document}